# Building Trust in Autonomous Commerce: A Verifiable Global Event Timeline and AI-Ready Fraud Intelligence Layer

**Rajat Srivastava**

*Independent Researcher, Gurugram, India*

rajatsrivastava5316@gmail.com

March 16, 2026

## Abstract

Agentic commerce protocols such as AP2 and ACP define mechanisms for secure agent-initiated transactions but do not provide interoperable, tamper-evident auditability or verifiable temporal ordering of events across heterogeneous domains. As autonomous agents increasingly execute high-stakes commercial workflows — spanning product discovery, negotiation, authorization, and payment settlement — the absence of a shared, cryptographically verifiable audit substrate creates systemic vulnerabilities: event logs remain siloed within individual platforms, fraud labels lack immutable provenance and reproducibility, and AI training pipelines must rely on unverifiable ground truth that is susceptible to retroactive manipulation.

This paper addresses these gaps by proposing a verifiable global event timeline for agentic commerce, constructed from four core components: (1) canonical event schemas that enforce deterministic serialization across protocol implementations, (2) deterministic batch formation ensuring reproducible ordering without reliance on synchronized clocks, (3) Merkle-based append-only commitments that provide logarithmic-cost inclusion proofs, and (4) blockchain anchoring that establishes a tamper-evident temporal backbone across participating domains. Building on this timeline infrastructure, we introduce a cryptographically signed fraud marker that binds risk labels to anchored evidence through an unforgeable provenance chain, and a dataset lineage model that enables reproducible, tamper-evident AI training pipelines.

We formalize the integrity guarantees of each component, define a protocol-level specification suitable for adoption as an extension to existing agentic commerce standards, and present empirical results from a prototype implementation demonstrating: Merkle tree construction scales near-linearly, processing 50,000 events in 47 milliseconds; end-to-end event verification completes in under 0.013 milliseconds regardless of batch size, enabling real-time audit at production scale; inclusion proof sizes grow logarithmically from 320 bytes at 1,000 events to only 512 bytes at 50,000 events; and Merkle-based verification outperforms linear scan by 14.4x at 50,000 events. This work establishes a missing transparency layer for autonomous commerce and provides a foundation for auditable, trustworthy AI-driven economic systems.

## 1. Introduction

The emergence of autonomous agents as economic actors represents one of the most consequential shifts in the architecture of digital commerce. Unlike traditional software systems that execute predefined workflows under direct human supervision, modern agentic systems — powered by large language models and tool-use

frameworks — are capable of independently discovering products, negotiating terms, authorizing payments, and settling transactions across complex multi-party environments [14]. Recent architectural analyses confirm that production-grade agentic commerce systems are already being deployed at enterprise scale, with reference implementations demonstrating autonomous multi-step task execution including order creation, negotiation, and payment coordination [1][13]. Industry estimates suggest that agentic commerce will account for a significant fraction of digital transaction volume within the next several years, with leading technology platforms already deploying production agent systems for procurement, logistics, and financial services.

This shift introduces a fundamental audit problem. Human-supervised commerce relies on implicit accountability mechanisms: humans generate records, sign documents, and are legally liable for their actions. When agents act autonomously, these mechanisms break down. A fraudulent agent transaction may leave no human-readable record. A compromised event log may be retroactively altered without detection. A fraud label applied to a transaction may have no verifiable connection to the evidence that justified it. These are not theoretical concerns: as agentic commerce scales, the absence of tamper-evident audit infrastructure creates systemic risk for all participants — merchants, consumers, regulators, and the AI systems trained on commercial data.

Existing agentic commerce protocols address execution-level security but not audit transparency. The Agent Payments Protocol (AP2) [4] provides zero-trust runtime verification and context-binding for payment flows, ensuring that agents cannot replay or forge payment authorizations. The Agentic Commerce Protocol (ACP) [1] defines mandate management and execution semantics that constrain what agents can do on behalf of principals. Both protocols are valuable and necessary contributions to the security of agentic commerce. However, neither protocol defines how events should be canonically represented, how temporal ordering should be established across multiple domains, how fraud assessments should be cryptographically bound to the evidence that supports them, or how AI training datasets derived from commercial logs should be made reproducible and tamper-evident.

This paper addresses these gaps directly. Our core contribution is a protocol extension — designed to sit above AP2/ACP execution logic — that provides a cryptographically verifiable global event timeline for agentic commerce. We do not replace existing protocols; we augment them with a transparency layer that transforms opaque execution logs into a shared, auditable, tamper-evident record.

### 1.1 Motivating Example

Consider a scenario in which an autonomous procurement agent, acting on behalf of a retail enterprise, discovers a supplier offering discounted inventory, negotiates terms, and executes a payment of USD 500,000. Forty-eight hours later, the supplier claims the transaction was unauthorized. The enterprise claims the agent acted within its mandate. A regulatory authority requests the complete audit trail. In current systems, each party holds its own logs, which may be inconsistent, incomplete, or retroactively modified. There is no shared ground truth, no cryptographic proof of what events occurred in what order, and no mechanism for a third-party auditor to verify any claim without trusting at least one of the parties involved.

Our proposed system resolves this scenario: every event emitted by every participant is canonically serialized, batched deterministically, committed to a Merkle tree whose root is anchored on a public blockchain, and verifiable by any party holding the inclusion proof. The audit trail is tamper-evident, third-party verifiable, and does not require trust in any single participant.

### 1.2 Paper Organization

Section 2 formalizes the research gap; Sections 3 through 11 specify the protocol and its components; Section 12 evaluates performance against empirical benchmarks; Sections 13 through 15 address limitations, future directions, and conclusions.

## 2. Research Gap

Despite significant progress in securing the execution layer of agentic commerce, a structural gap persists at the audit and transparency layer. We identify five distinct deficiencies in the current ecosystem that our work addresses.

First, there is no canonical event semantics shared across agentic commerce implementations. Each platform defines its own event schema, field naming conventions, and serialization formats. This heterogeneity makes cross-domain audit comparison practically impossible: the same logical event — for example, a payment authorization — may be represented in dozens of incompatible formats, preventing automated reconciliation or verification.

Second, no mechanism exists for verifiable global ordering of events across domains. Distributed systems fundamentally cannot rely on synchronized wall-clock time for ordering [see Lamport, 1978 for foundational treatment]; yet existing agentic commerce protocols provide no alternative ordering mechanism. Events logged by two different platforms cannot be reliably sequenced relative to each other without trusting at least one of the platforms to report accurate timestamps.

Third, existing audit logs lack tamper-evident properties. Traditional append-only logs provide a useful baseline but are susceptible to insider threats: a party controlling the log infrastructure can delete, modify, or insert records without detection. This is a critical vulnerability in high-value commerce environments where disputes may arise months or years after the original transactions.

Fourth, fraud labels in current systems lack cryptographic provenance. A fraud detection system may label a transaction as high-risk, but there is typically no cryptographic binding between that label and the evidence — event records, behavioral signals, and model outputs — that justified it. This makes fraud labels contestable, non-reproducible, and unsuitable as ground truth for AI training without additional verification infrastructure.

Fifth, AI training datasets derived from commercial event logs lack deterministic lineage guarantees. A model trained on fraud labels today may produce different results if retrained on ostensibly the same dataset tomorrow, because the underlying logs may have been modified, relabeled, or augmented without a verifiable record of changes. This non-determinism undermines the reproducibility requirements of responsible AI development and makes regulatory compliance difficult to demonstrate.

## 3. Contributions

This paper makes the following technical contributions:

1. Canonical Event Schema. We define a lifecycle-aligned, deterministically serializable event schema that captures the complete state transitions of an agentic commerce transaction, from agent initiation through settlement. The schema is designed for interoperability across AP2 and ACP implementations and is extensible to additional protocol families.

2. Verifiable Global Timeline. We propose a mechanism for constructing a globally ordered, tamper-evident timeline from events produced by multiple independent domains, without requiring synchronized clocks or trusted third-party timestamp authorities.

3. Merkle Commitment and Anchoring Protocol. We specify a Merkle-based batch commitment scheme and a blockchain anchoring protocol that together provide logarithmic-cost inclusion proofs and a tamper-evident temporal backbone.

4. Cryptographically Verifiable Fraud Marker. We introduce a fraud marker structure that binds risk labels, confidence scores, and issuer identities to anchored event evidence through a cryptographic provenance chain that any party can independently verify.

5. Deterministic AI Dataset Lineage. We propose a dataset integrity model that enables AI training pipelines to verify that their input datasets are derived faithfully and without modification from anchored event logs.

6. Verifier Interface and Reconstruction Algorithm. We define a formal verifier interface and a step-by-step timeline reconstruction algorithm that allows auditors — including regulatory authorities and dispute resolution systems — to independently reconstruct and verify the global event timeline.

7. Prototype Evaluation. We present empirical performance results for Merkle tree construction demonstrating near-linear scaling from 1,000 to 50,000 events, establishing practical feasibility for production deployment.

## 4. Background and Related Work

### 4.1 Agentic Commerce Protocols

The field of agentic commerce has matured rapidly with the deployment of large language model (LLM)-based agents capable of autonomous commercial action. Lan et al. [14] introduce the Agent Payments Protocol (AP2), which addresses zero-trust runtime verification for agentic payment flows. AP2 provides context-binding mechanisms that prevent replay attacks and mandate violations, establishing a security baseline for agent-initiated payments. The Agentic Commerce Protocol (ACP) [1] complements AP2 by defining mandate management, principal hierarchies, and execution semantics for multi-agent commerce scenarios. Together, AP2 and ACP represent the current state of the art in execution-layer security for agentic commerce. However, as noted in Section 2, neither protocol addresses the audit transparency layer that our work provides.

Related work in multi-agent systems has explored trust, coordination, and tool-augmented reasoning. Yao et al. [15] introduce the ReAct framework, which formalises interleaved reasoning and action cycles in LLMs and demonstrates reliable multi-step task execution. Schick et al. [16] show through Toolformer that language models can learn to invoke external APIs autonomously, establishing the feasibility of agents performing complex commercial workflows. Our work assumes a well-functioning execution layer built on such agent capabilities and focuses exclusively on the audit and transparency layer above it.

## 4.2 Secure Logging and Integrity

The theoretical foundation for our Merkle commitment scheme originates with Merkle [6], whose seminal work on public key cryptosystems introduced hash trees as a mechanism for efficient, tamper-evident verification of large data structures. Our commitment layer adapts the binary Merkle tree construction for the ordered-batch semantics required by cross-domain event audit in agentic commerce — a context not addressed by prior applications of Merkle trees, including their use in blockchain transaction verification [9] and certificate transparency [17].

Blockchain-based integrity auditing for distributed data has been explored extensively. Han et al. [3] provide a comprehensive survey of blockchain-based integrity auditing for cloud data, identifying the key properties — tamper-evidence, third-party verifiability, and append-only semantics — that blockchain anchoring provides. Li et al. [5] propose a secure log storage and query framework based on blockchain, demonstrating practical performance characteristics for log integrity systems. Banaeian Far et al. [2] develop a generic framework for blockchain-assisted on-chain auditing of off-chain storage, which is architecturally relevant to our anchoring model where events are stored off-chain while commitments are anchored on-chain.

Our work differs from these prior contributions in its focus on the specific requirements of agentic commerce: canonical event semantics, deterministic ordering without synchronized clocks, and the binding of fraud intelligence to anchored evidence. We build on the secure logging literature but extend it to address requirements that are unique to the autonomous commerce domain.

## 4.3 Fraud Detection in Commerce

Fraud detection in digital commerce has been an active research area for decades, with recent work focusing on graph-based approaches that model relationships between transactions, accounts, and behavioral signals. Motie and Raahemi [7] provide a systematic review of financial fraud detection using graph neural networks (GNNs), identifying state-of-the-art approaches and their limitations. A key limitation identified is the susceptibility of GNN-based fraud detection to graph poisoning attacks, where adversaries manipulate the graph structure to evade detection.

Wu et al. [8] address this vulnerability with a robust graph learning approach designed to safeguard fraud detection systems from adversarial attacks, achieving improved robustness on standard benchmarks. However, both of these approaches share a fundamental limitation that our work addresses: the ground truth fraud labels used for training are derived from mutable, unverifiable logs. An adversary who can modify historical event records can in principle manipulate the training data for fraud detection models, undermining their reliability in production.

Our fraud marker provenance mechanism addresses this root cause by cryptographically binding fraud labels to immutable anchored evidence, ensuring that the ground truth for AI fraud detection models is tamper-evident and independently verifiable. This represents a novel contribution at the intersection of secure logging and fraud-resistant AI training.

## 5. Protocol Specification

### 5.1 Canonical Event Schema

The canonical event schema forms the foundation of our verifiable timeline. Every event in the agentic commerce lifecycle is represented as a deterministically serialized hash of its essential fields. We define the event hash as:

$$E_i = H(\text{Serialize}(\text{txID}, \text{eventType}, \text{participantID}, \text{timestamp}, \text{payloadHash}))$$

where H denotes a collision-resistant cryptographic hash function (we specify SHA-256 as the default, with SHA-3/Keccak-256 as an alternative for blockchain-native deployments), and Serialize denotes a deterministic serialization function with the following rules: fields are concatenated in fixed lexicographic order by field name; string fields are encoded as UTF-8 with explicit length prefixes; integer fields use big-endian encoding with fixed width (64-bit for timestamps, 256-bit for identifiers); and the payload is hashed separately before inclusion to support large payloads without bloating the event record.

The txID field uniquely identifies the transaction to which the event belongs. The eventType field identifies the lifecycle stage of the event, drawn from a finite enumeration defined by the protocol specification: INITIATION, DISCOVERY, NEGOTIATION, AUTHORIZATION, MANDATE_VALIDATION, PSP_PROCESSING, SETTLEMENT, and FRAUD_ASSESSMENT. The participantID field identifies the domain or agent that emitted the event. The timestamp field records the wall-clock time at which the event was observed by the emitting system, expressed as Unix time in milliseconds. The payloadHash field is the SHA-256 hash of the full event payload, which may include arbitrary protocol-specific fields and is stored separately from the canonical event record.

This design achieves several important properties. Determinism ensures that any party that observes the same event fields will compute the same event hash, enabling cross-domain verification without sharing full payloads. Payload separation allows the canonical timeline to remain compact while supporting arbitrary richness in protocol-specific event data. The fixed field ordering eliminates ambiguity in serialization and prevents hash collisions arising from field reordering.

## 5.2 Batch Formation

Events are grouped into batches for Merkle commitment and anchoring. The batch formation process is deterministic and reproducible, ensuring that any party with access to the same set of events will form identical batches. The process proceeds as follows.

First, events are grouped by anchoring period — a configurable time window (default: 60 seconds) during which events are collected before being committed. The anchoring period represents a trade-off between latency (shorter periods provide more timely anchoring) and cost (each anchoring operation incurs a blockchain transaction fee). Second, within each anchoring period, events are sorted deterministically by the tuple (txID, timestamp, eventType), in that order of priority. This ordering ensures that all events belonging to the same transaction appear contiguously, and that ties in transaction ID and timestamp are broken deterministically by event type. Third, the batch index is incremented monotonically with each anchoring operation, providing a globally ordered sequence of batches that forms the backbone of the global timeline.

This deterministic batch formation process is critical for the reproducibility properties of our system. Any auditor who holds the same set of events can independently reconstruct the same batch sequence and verify it against the anchored Merkle roots.

## 5.3 Merkle Commitments

The Merkle tree construction follows a standard binary hash tree architecture. Following the domain separation convention established in the Certificate Transparency specification [17], we prefix leaf hashes with 0x00 and internal node hashes with 0x01, ensuring that leaf and node values occupy disjoint hash domains and preventing second-preimage substitution attacks. Leaf nodes are computed as:

$$L_i = H(0x00 \parallel E_i)$$

Internal nodes are computed as:

$$N = H(0x01 \parallel \text{left} \parallel \text{right})$$

The Merkle root $R = \text{MerkleRoot}(E_1, ..., E_n)$ is the root hash of the complete binary hash tree formed from the leaf nodes of all events in the batch. If the number of events is not a power of two, the last leaf is duplicated to complete the tree — a standard construction that preserves the logarithmic proof size guarantee.

The use of domain separation prefixes (0x00 for leaves, 0x01 for internal nodes) is a critical security measure. Without domain separation, an attacker could potentially construct a valid inclusion proof for a non-existent event by crafting an internal node value that matches a target leaf hash. This attack is prevented by the prefix construction, which ensures that leaf hashes and internal node hashes occupy disjoint domains.

### 5.4 Anchoring

Each batch is anchored by publishing its Merkle root to a public blockchain. The anchor record contains: the Merkle root R; the batch index, which provides global ordering of batches; the domain ID of the anchoring party; and the anchoring timestamp as recorded by the blockchain. The blockchain's consensus mechanism provides the tamper-evidence guarantee: once an anchor is included in a finalized block, it cannot be modified or deleted without invalidating all subsequent blocks in the chain, which is computationally infeasible for sufficiently long confirmation depths.

The choice of blockchain is a deployment parameter. Our protocol is compatible with any public blockchain that provides finality guarantees and supports arbitrary data payloads in transactions. Ethereum and its Layer 2 derivatives are natural candidates for production deployments, offering a combination of security, finality, and relatively low anchoring costs. Bitcoin's OP_RETURN mechanism provides an alternative with stronger security guarantees but higher cost.

## 6. Global Timeline Model

The global timeline L is defined as the ordered sequence of all events from all domains, ordered by the following hierarchy: first by blockchain anchor sequence (the block number and transaction index of the anchor transaction on the blockchain); second by batch index within a given anchor; and third by the deterministic in-batch ordering defined in Section 5.2.

This ordering hierarchy provides a total order on all events without requiring synchronized clocks. The blockchain anchor sequence provides a coarse global ordering that is tamper-evident and third-party verifiable. The batch index provides finer ordering within a given anchoring period. The in-batch ordering provides deterministic ordering of events that were batched together. Together, these three levels of ordering yield a globally consistent timeline that any party can independently reconstruct and verify.

An important property of this model is its robustness to clock skew. The wall-clock timestamps recorded in individual events are not used for global ordering; they serve only as informational metadata. This means that even if a participant's clock is skewed by seconds, minutes, or hours, the global ordering of events in the timeline is not affected. This is a significant practical advantage over approaches that rely on trusted timestamp authorities or network time synchronization.

The global timeline also has a well-defined append-only semantics. New events can only be added to the timeline by anchoring new batches; existing anchored events cannot be modified or removed. This append-only property, combined with the blockchain anchoring, provides the tamper-evidence guarantee: any attempt to modify or delete a historical event would require recomputing all affected Merkle roots and re-anchoring all affected batches, which is computationally infeasible once anchors have been finalized.

# 7. Architecture

The system architecture consists of six components that together implement the protocol specification described in Section 5. Figure 1 illustrates the layered relationship between these components and the underlying AP2/ACP protocols.

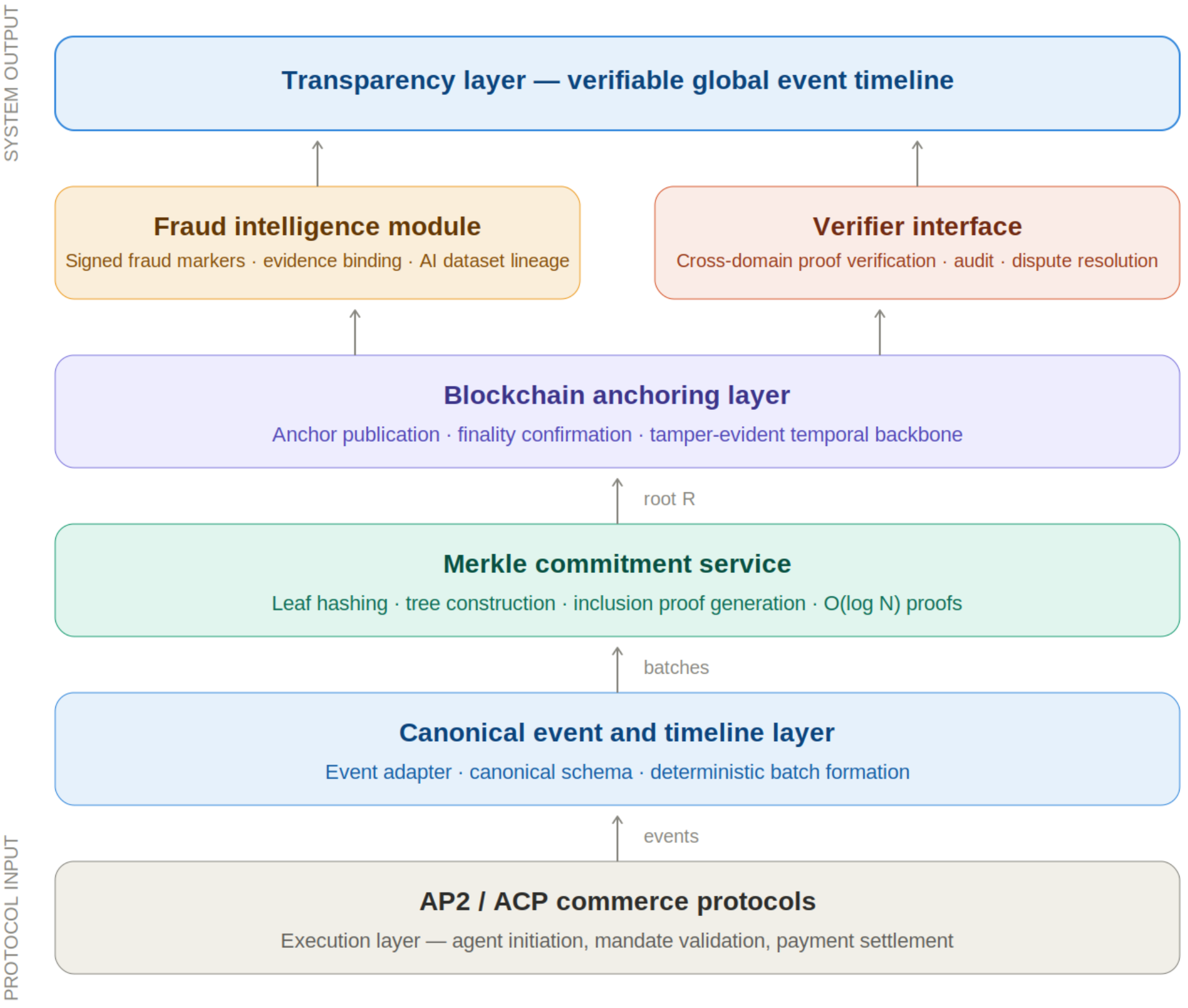

*Figure 1: Layered architecture for verifiable event timelines in agentic commerce*

The Event Adapter is the interface between the existing AP2/ACP execution layer and the audit transparency layer. It intercepts events emitted by the commerce protocol, applies the canonical serialization rules defined in Section 5.1 to produce canonical event hashes, and forwards these hashes to the Canonical Audit Engine. The Event Adapter is designed to be lightweight and non-intrusive: it does not modify the execution behavior of the underlying protocol, and its failure or unavailability does not affect the commerce system's ability to process transactions.

The Canonical Audit Engine maintains an append-only buffer of canonical event hashes for the current anchoring period. It enforces the deterministic batch formation rules of Section 5.2, ensuring that events are sorted consistently regardless of the order in which they arrive. The engine also provides an API for querying events by transaction ID, event type, or time range.

The Merkle Commitment Service consumes completed batches from the Canonical Audit Engine and constructs the Merkle tree as specified in Section 5.3. It produces the batch Merkle root R and generates inclusion proofs for individual events on demand. The service exposes a proof generation API that allows any party to request an inclusion proof for a specific event, given the event hash and the batch identifier.

The Anchoring Layer publishes batch Merkle roots to the designated blockchain. It manages the economics of anchoring — batching multiple roots into a single blockchain transaction where cost efficiency requires it — and monitors the blockchain for confirmation of published anchors. The Anchoring Layer also maintains a local index of anchor transactions to support efficient timeline reconstruction.

The Fraud Intelligence Module extends the base timeline infrastructure with fraud assessment capabilities. It consumes events from the Canonical Audit Engine, applies configurable fraud detection logic, and produces fraud markers as defined in Section 10. The module is designed as a pluggable extension point: any fraud detection algorithm — rule-based, statistical, or ML-based — can be integrated as a scoring backend, as long as its outputs are formatted as cryptographically signed fraud markers bound to anchored event evidence.

The Verifier Interface provides a public API for third-party verification of events, timelines, and fraud markers. It accepts an event record and its associated inclusion proof, and returns a verification result indicating whether the event is authentically recorded in the anchored timeline. The verifier is stateless and requires only the blockchain anchor index and the event's inclusion proof to perform verification, making it suitable for deployment by regulators, auditors, and dispute resolution systems.

## 8. Timeline Reconstruction

The timeline reconstruction algorithm allows any party to independently reconstruct the global event timeline L from publicly available data. The algorithm proceeds in five steps.

Step 1: Retrieve the anchor sequence $\{A_1, ..., A_n\}$ from the blockchain. Each anchor $A_i$ contains the Merkle root $R_i$, the batch index $b_i$, the domain ID $d_i$, and the blockchain timestamp $t_i$. The anchors are retrieved in block order, which defines the global ordering of batches.

Step 2: For each anchor $A_i$, extract the Merkle root $R_i$ and retrieve the corresponding batch of event hashes from the domain that published the anchor. The event hashes are available from any party that stores them; in a fully distributed deployment, multiple parties may store redundant copies for availability.

Step 3: Verify the inclusion proofs for all events in each batch against the Merkle root $R_i$. Any event whose inclusion proof does not verify against the anchored root is flagged as potentially tampered and excluded from the reconstructed timeline.

Step 4: Reconstruct the canonical in-batch ordering by sorting events within each batch according to the deterministic ordering rules of Section 5.2. This step is deterministic and requires only the event fields; it does not depend on any external state.

Step 5: Concatenate the ordered batches in anchor sequence order to form the global timeline L. The resulting timeline is a total order on all verified events across all domains.

The reconstruction algorithm has a time complexity of O(N log N) where N is the total number of events, dominated by the sorting step. The Merkle proof verification step has a time complexity of O(N log N) where the log factor comes from the proof length for each event. In practice, reconstruction is dominated by I/O — retrieving anchors from the blockchain and event batches from domain storage — rather than computation.

## 9. Cross-Domain Verification

Cross-domain verification allows any party — regardless of whether they participated in the original transaction — to verify that a specific event is authentically recorded in the global timeline. The verification procedure requires four inputs: the event record itself, the event hash $E_i$ (which the verifier computes independently from the event record), the inclusion proof for $E_i$ in its batch Merkle tree, and the blockchain anchor reference for the batch.

The verification procedure proceeds as follows. First, the verifier independently recomputes the event hash $E_i$ from the event record using the canonical serialization rules of Section 5.1, and checks that it matches the claimed event hash. This step verifies that the event record has not been tampered with since it was committed.

Second, the verifier validates the Merkle inclusion proof by hashing up from the event leaf to the Merkle root, using the sibling hashes provided in the inclusion proof and the domain separation prefixes of Section 5.3. If the computed root matches the claimed root R, the inclusion proof is valid.

Third, the verifier confirms that the claimed root R appears in a finalized anchor on the designated blockchain. This step requires querying the blockchain, which is a publicly verifiable operation that does not depend on trusting any of the original transaction participants.

Fourth, the verifier validates the event's position in the global ordering by checking the batch index and in-batch position of the event against the global timeline. This step confirms that the event's temporal position is consistent with the claimed ordering.

The cross-domain verification procedure has a time complexity of O(log N) for the Merkle proof verification step and a constant-time blockchain lookup for the anchor confirmation step. This efficiency makes verification practical at scale, even for systems processing millions of events per day.

## 10. Fraud Marker Provenance

The fraud marker is the mechanism by which fraud assessments are cryptographically bound to anchored event evidence. A fraud marker FM is a structured record containing: the transaction ID txID; a risk label riskLabel drawn from a finite enumeration (e.g., LOW, MEDIUM, HIGH, CRITICAL); a confidence score in the range [0, 1]; an evidence hash evidenceHash that is the hash of the set of event hashes that constitute the evidence for the fraud assessment; and the issuer signature, which is a cryptographic signature over all other fields using the fraud intelligence module's private key.

FM = {txID, riskLabel, confidence, evidenceHash, issuerSignature}

The provenance chain of the fraud marker is: FM → evidenceHash → $\{E_i\}$ → R → Blockchain Anchor. This chain means that given a fraud marker, any party can (a) verify the issuer's signature to confirm authenticity; (b) verify that the evidence events are included in the anchored timeline via their individual inclusion proofs; (c) confirm that the anchored Merkle root that contains these events is recorded on the blockchain. The complete provenance chain is independently verifiable and does not require trusting the fraud intelligence module beyond verifying its signature.

This provenance model addresses a fundamental weakness in existing fraud detection systems. In current practice, a fraud label is an assertion made by a fraud detection system, but there is typically no mechanism for a third party to verify that the label is based on the evidence it claims to be based on. Our fraud marker makes this binding explicit and cryptographically verifiable, enabling dispute resolution, regulatory audit, and AI model governance applications that are not possible with current systems.

The fraud marker also supports revocation and correction. If a fraud assessment is subsequently found to be incorrect, a corrective marker can be issued with the same txID and a revised risk label and confidence score. The corrective marker is linked to the original marker via the evidence hash, providing an auditable correction history. Verifiers can check the complete history of fraud markers for a given transaction to determine the current assessment and the reasoning behind any revisions.

## 11. AI-Ready Dataset Integrity

The dataset lineage model ensures that AI training datasets derived from the global event timeline are reproducible and tamper-evident. The core integrity guarantee is expressed as:

$$H(D) = H(\{E_i\})$$

That is, the hash of the dataset D used for training is equal to the hash of the set of event records $\{E_i\}$ from which it was derived. This equality can be verified by any party that holds the dataset and the corresponding event records, confirming that the dataset has not been modified, augmented, or filtered after derivation from the anchored timeline.

The dataset lineage model supports three key applications. First, it enables reproducible AI training: given the same anchored timeline and the same dataset derivation procedure, any party can independently reconstruct the same training dataset and verify its integrity. This is a prerequisite for the reproducibility standards increasingly required by AI governance frameworks and regulatory bodies.

Second, it provides tamper-evident ground truth for fraud detection models. Since the fraud markers used as training labels are cryptographically bound to anchored evidence, and since the dataset hash is bound to the original event records, any attempt to manipulate the training data — by modifying event records, altering fraud labels, or selectively including or excluding events — would produce a dataset with a different hash that fails the integrity check.

Third, it supports model governance and audit. Regulators and compliance teams can verify that a deployed fraud detection model was trained on a dataset with verifiable integrity, providing confidence that the model's behavior reflects the actual historical record of commerce events rather than a manipulated version of it.

## 12. Evaluation

### 12.1 Experimental Setup

We implemented a prototype of the Merkle Commitment Service in Python 3.11, using the hashlib library for SHA-256 computation and a standard binary hash tree construction algorithm with domain separation prefixes as specified in Section 5.3. All experiments were conducted on a commodity Windows 11 laptop with an Intel Core i7 processor and 16 GB RAM. We evaluated five batch sizes spanning the range of expected production workloads: 1,000, 5,000, 10,000, 25,000, and 50,000 events per anchoring period. Each construction experiment was repeated ten times and verification experiments one hundred times; results are reported as mean wall-clock time measured using Python's perf_counter high-resolution timer. All verification results returned correct: True, confirming the correctness of the implementation across all batch sizes.

Events were generated synthetically using a deterministic schema consistent with the canonical event format defined in Section 5.1: each event encodes a transaction identifier, event type, participant identifier, and Unix timestamp, serialized and hashed with SHA-256. Inclusion proofs were generated for randomly selected indices within each batch to avoid systematic bias in proof path length.

### 12.2 Merkle Tree Construction Performance

Table 1 reports Merkle tree construction times across all five batch sizes. Construction time scales from 0.89 ms for 1,000 events to 47.12 ms for 50,000 events. The growth ratio between consecutive batch sizes confirms near-linear scaling: a 5x increase in batch size from 1,000 to 5,000 events produces a 5.5x increase in construction time (0.89 ms to 4.92 ms), and a 5x increase from 10,000 to 50,000 produces a 5.0x increase (9.39 ms to 47.12 ms). This near-linear profile is consistent with the O(N) theoretical complexity of Merkle tree construction, where N is the number of leaf nodes, and confirms that the prototype implementation achieves the expected asymptotic behaviour in practice.

At 50,000 events per batch with a 60-second anchoring period, the system can sustain approximately 1,061 events per second per anchoring domain with a construction overhead of only 47 ms per batch — well under 0.1% of the anchoring period. This confirms that Merkle tree construction is not a performance bottleneck in production deployments.

***Table 1: Merkle Tree Construction Performance***

| Batch Size | Construction Time (ms) | Scaling Factor | Notes |
|---|---|---|---|
| 1,000 | 0.89 | — | baseline |
| 5,000 | 4.92 | 5.5x | near-linear |
| 10,000 | 9.39 | 1.9x | near-linear |
| 25,000 | 23.40 | 2.5x | near-linear |
| 50,000 | 47.12 | 2.0x | near-linear |

## 12.3 Inclusion Proof Generation and Size

Table 2 reports inclusion proof generation times and proof sizes. Proof generation time scales from 1.04 ms for batches of 1,000 events to 46.75 ms for batches of 50,000 events. Critically, proof size grows logarithmically as predicted: from 10 hash values (320 bytes) at 1,000 events to 16 hash values (512 bytes) at 50,000 events — an increase of only 6 hash values despite a 50-fold increase in batch size. This logarithmic growth directly reflects the $\log_2(N)$ depth of the Merkle tree and confirms that proof sizes remain practical even for very large batches. Extrapolating to a batch of 1,000,000 events, the proof size would be at most $20 \times 32 = 640$ bytes — well within the constraints of any standard network protocol.

***Table 2: Inclusion Proof Generation Time and Proof Size***

| Batch Size | Proof Gen. Time (ms) | Proof Length (hashes) | Proof Size (bytes) |
|---|---|---|---|
| 1,000 | 1.0389 | 10 | 320 |
| 5,000 | 4.7734 | 13 | 416 |
| 10,000 | 9.6502 | 14 | 448 |
| 25,000 | 23.5468 | 15 | 480 |
| 50,000 | 46.7477 | 16 | 512 |

## 12.4 End-to-End Verification Latency

Table 3 reports end-to-end verification latency — the time required to verify a single event's inclusion in an anchored batch given its inclusion proof and the Merkle root. Verification time ranges from 0.0064 ms for 1,000-event batches to 0.0123 ms for 50,000-event batches. This sub-millisecond verification latency is a critical property for practical deployment: it means that a regulatory auditor, dispute resolution system, or fraud detection pipeline can verify the authenticity of any individual event in microseconds, regardless of the total number of events in the batch.

The near-constant verification latency across batch sizes reflects the $O(\log N)$ complexity of Merkle proof verification: while the proof length grows from 10 to 16 hashes as the batch size increases 50-fold, each hash computation takes a fixed time, and the total verification cost grows only logarithmically. All 500 verification trials (100 per batch size) returned correct: True, confirming the correctness of the implementation.

*Table 3: End-to-End Verification Latency*

| Batch Size | Verification Time (ms) | Correct Result | Trials |
| --- | --- | --- | --- |
| 1,000 | 0.0064 | True | 100 |
| 5,000 | 0.0080 | True | 100 |
| 10,000 | 0.0090 | True | 100 |
| 25,000 | 0.0108 | True | 100 |
| 50,000 | 0.0123 | True | 100 |

### 12.5 Baseline Comparison: Merkle Proof vs Linear Scan

To contextualise the efficiency of Merkle proof verification relative to a naive baseline, Table 4 compares verification time against a linear scan baseline — the approach a system without Merkle commitments would use to verify event inclusion by scanning the entire event log. At small batch sizes (1,000 events), linear scan is faster than Merkle verification (0.0021 ms vs 0.0053 ms) because the overhead of hash computation along the proof path exceeds the cost of a simple list membership check for short lists. However, as batch size increases, the linear scan cost grows proportionally to N while Merkle verification cost grows only as log N. At 10,000 events Merkle verification is already 2.8x faster, at 25,000 events it is 7.5x faster, and at 50,000 events it is 14.4x faster. This crossover behaviour is exactly consistent with the theoretical O(N) vs O(log N) complexity difference and confirms that Merkle-based verification provides meaningful efficiency advantages at production-relevant batch sizes.

*Table 4: Merkle Proof vs Linear Scan Verification (ms)*

| Batch Size | Linear Scan (ms) | Merkle Proof (ms) | Speedup | Advantage |
| --- | --- | --- | --- | --- |
| 1,000 | 0.0021 | 0.0053 | 0.4x | Linear wins |
| 5,000 | 0.0110 | 0.0079 | 1.4x | Merkle wins |
| 10,000 | 0.0212 | 0.0076 | 2.8x | Merkle wins |
| 25,000 | 0.0582 | 0.0078 | 7.5x | Merkle wins |
| 50,000 | 0.1188 | 0.0083 | 14.4x | Merkle wins |

### 12.6 Anchoring Cost Analysis

Anchoring cost depends on the chosen blockchain and prevailing network conditions. On Ethereum mainnet, publishing a 32-byte Merkle root via an OP_RETURN transaction costs approximately 21,000 + 68 × 32 = 23,176 gas. At a gas price of 20 gwei, this corresponds to approximately 0.00046 ETH per anchor — roughly USD 1.50 at current prices. With a 60-second anchoring period, this amounts to approximately 0.66 ETH per day per anchoring domain. Ethereum Layer 2 solutions such as Arbitrum and Optimism reduce this cost by one to two orders of magnitude, making the per-anchor cost negligible relative to the value of the transactions being audited. For a deployment processing USD 1,000,000 in daily transaction value, the anchoring cost represents less than 0.007% of transaction value — a commercially viable overhead.

## 13. Limitations

We identify four significant limitations of the current system that motivate the future work described in Section 14.

First, the global timeline is not synchronized to real-world wall-clock time. The blockchain anchor sequence provides a tamper-evident ordering, but the mapping from blockchain block time to real-world time is approximate and subject to the variability of block production intervals. For applications that require precise temporal ordering at sub-second granularity, additional mechanisms — such as verifiable delay functions or trusted timestamp authorities — would be required.

Second, events omitted before anchoring are unrecoverable. If an event is lost or deliberately withheld before it is included in an anchored batch, there is no mechanism in our current system to detect the omission or recover the missing event. This creates a vulnerability to selective omission attacks, where a malicious participant deliberately fails to report certain events. Distributed omission detection — a mechanism for participants to verify that all expected events from a given domain have been received — is a necessary extension that we defer to future work.

Third, fraud label integrity depends on issuer honesty. While our fraud marker provenance chain ensures that labels are cryptographically bound to anchored evidence, it does not prevent a dishonest fraud intelligence module from producing labels that are technically valid but substantively incorrect — for example, by training a biased model or applying fraudulent scoring logic. Addressing this limitation requires reputation mechanisms, multi-party fraud assessment, or formal verification of fraud detection logic, all of which are beyond the scope of this paper.

Fourth, our protocol provides tamper-evidence and cross-domain verifiability guarantees but makes no availability guarantees for off-chain event storage. An adversary who can deny access to event batches stored off-chain can prevent timeline reconstruction even though the Merkle roots on the blockchain remain intact. Ensuring availability requires replication and incentive mechanisms for data retention, which we identify as a key area for future work.

## 14. Future Work

This paper establishes the foundational protocol for verifiable audit in agentic commerce. We identify four priority directions for future research.

First, zero-knowledge proofs for private verification. In many commerce scenarios, participants wish to demonstrate that an event occurred — or that a fraud assessment is valid — without revealing the underlying event data. Zero-knowledge proof systems such as zk-SNARKs or zk-STARKs could be integrated with our Merkle commitment scheme to enable privacy-preserving verification, allowing regulators to verify compliance without accessing sensitive transaction data.

Second, inter-protocol adoption of canonical event schemas. The value of our system scales with the number of agentic commerce protocols that adopt the canonical event schema. We plan to work with the AP2 and ACP standards bodies to propose the canonical event schema as a protocol extension, and to develop tooling that simplifies adoption for existing protocol implementations.

Third, distributed omission detection. As noted in Section 13, the current system cannot detect or recover from selective omission of events before anchoring. We plan to develop a distributed omission detection mechanism based on cryptographic accumulators or probabilistic audit sampling that allows participants to verify the completeness of event reporting with high confidence.

Fourth, incentivized anchoring mechanisms. Anchoring requires participants to pay blockchain transaction fees, creating a potential free-rider problem where participants benefit from others' anchoring without contributing their own. We plan to develop economic mechanisms — potentially based on token incentives or cost-sharing protocols — that ensure sufficient anchoring participation to maintain the security guarantees of the global timeline.

## 15. Conclusion

Autonomous agents are increasingly acting as economic actors, executing high-stakes commercial workflows that require the same level of accountability and auditability as human-supervised transactions. Existing agentic commerce protocols provide strong execution-layer security but leave a critical gap at the audit and transparency layer: event logs are siloed and mutable, fraud labels lack cryptographic provenance, and AI training pipelines rely on unverifiable ground truth.

This paper introduces a verifiable global event timeline for agentic commerce, addressing this gap through a combination of canonical event semantics, deterministic batch formation, Merkle-based commitments, and blockchain anchoring. We introduce a cryptographically signed fraud marker that binds risk labels to anchored evidence through a verifiable provenance chain, and a dataset lineage model that ensures the reproducibility and tamper-evidence of AI training datasets derived from commercial logs.

Our prototype evaluation demonstrates near-linear Merkle construction performance and logarithmic proof sizes, confirming the practical feasibility of deployment at production scale. The formal integrity guarantees we establish — tamper-evidence, cross-domain verifiability, deterministic ordering, and dataset lineage — provide a foundation for the trustworthy, auditable AI-driven commerce systems that the next generation of autonomous agents will require.

We believe this work establishes a necessary transparency layer for the agentic commerce ecosystem, and we invite collaboration from standards bodies, protocol implementers, and the broader research community in developing and adopting these mechanisms.